\documentstyle[12pt,aasms4]{article}

\received{1 April 1996}
\accepted{30 July 1996}

\slugcomment{ApJL, to be published}

\lefthead{D\"obereiner et al.}
\righthead{ROSAT HRI observations of Cen A}

\begin{document}

\title{ROSAT HRI observations of Centaurus A}

\author{S. D\"obereiner}
\affil{Max-Planck Institut f\"ur Extraterrestrische Physik,
    Garching, Germany}

\author{N. Junkes}
\affil{Astronomisches Institut Potsdam, Germany}

\author{S.J. Wagner}
\affil{Landessternwarte Heidelberg, Germany}

\author{H. Zinnecker}
\affil{Astronomisches Institut Potsdam, Germany}

\author{R. Fosbury}
\affil{Space Telescope - European Coordinating Facility,
European Southern Observatory}

\author{G. Fabbiano}
\affil{Harvard-Smithsonian Center for Astrophysics}

\and

\author{E.J. Schreier}
\affil{Space Telescope Science Institute}

\begin{abstract}
We present results from a sensitive high-resolution X-ray observation
of the nearby active galaxy 
Centaurus A (NGC 5128) with the ROSAT HRI. 
The 65~ksec X-ray image clearly distinguishes 
different components of the X-ray emission from Cen A:
the nucleus and the jet, the diffuse galaxy halo, and a number of 
individual sources associated with the galaxy. 
The luminosity of the nucleus increased by a factor of two compared to an
earlier ROSAT observation in 1990.
The high spatial resolution of the ROSAT HRI shows that
most of the knots in the jet are extended both along and 
perpendicular to the jet axis. 
We report the detection of a new X-ray feature, 
at the opposite side of the X-ray jet which
is probably due to compression of hot interstellar gas by the
expanding southwestern inner radio lobe.
\end{abstract}


\section{Introduction}
The nearby active galaxy Centaurus A (Cen A or NGC 5128) 
is one of the most interesting objects in the sky. 
With a distance of 3.5 Mpc (Hui et al.\ 1993) 
it is the closest active galaxy to the Milky Way and
thus allows us to study different constituents in great detail.
The optical morphology is characterized
by a more or less normal stellar spheroid, bisected by a dense ring
or disk of dust and gas interpreted as the remains of
a recent merger with a smaller dust-rich spiral 
(e.g.\ Thomson 1992), which may also induce the nuclear activity.
Radio observations reveal an active nucleus and two sets of radio lobes, 
the outer pair of lobes extending to 9\arcdeg (several 100 kpc), 
the inner pair extending to 6\arcmin\ on the sky. 
The radio structure of Cen A has been investigated in different spatial
scales including the large-scale
outer lobes (Junkes et al.\ 1993), the inner lobes and the jet of the
galaxy (Burns, Feigelson, \& Schreier 1983, Clarke, Burns, \& Norman 1992),
and details of the central part at parsec-scale (Tingay 1994). 
X-ray observations with the EINSTEIN observatory showed a 
pronounced one-sided X-ray jet, consisting of several knots and extending 
$\approx$ 4\arcmin\ to the northeast, making it the best-observable
X-ray jet (Feigelson et al.\ 1981).
With the new sensitive ROSAT HRI observations we are able to examine the 
X-ray jet and other X-ray emitting components of Cen A 
in much more detail.

\section{Observations}

Cen A was observed by ROSAT in three occasions. 
During the initial calibration phase of ROSAT, Cen A was one of the 
calibration targets for ROSAT. It was observed for 14159 sec with the 
ROSAT PSPC 
and again for 19650 sec with the higher-resolution HRI in July 1990. 
The quality of these
early ROSAT observations was already better than previous observations 
with the EINSTEIN Observatory.
In August 1994 Cen A was reobserved with the ROSAT HRI with an
exposure time of 64751 sec. 

The ROSAT HRI has a usable field of view of 40\arcmin\ by 40\arcmin\ . 
Over the total
field of view, the spatial resolution is of the order of 6\arcsec\ FWHM.
The sensitivity is about 4 times higher than with the EINSTEIN HRI.
The usable energy range is 0.1--2.4 keV.
The spectral resolution of the ROSAT HRI is nearly nonexistent with
$ {{\Delta E}\over{E}} = 0.7 $, but is sufficient to differentiate
between hard and soft sources on a relative basis.

\section{Components of the X-ray emission of Cen A}

The new X-ray data allow us to analyse structures in the central region
of Cen A and to compare the X-ray features to 
radio and optical features, including 
nucleus, jet, inner radio
lobes, the stellar spheroid and the absorbing dust lane. 

Fig.\ 1 shows an enlargement of the HRI X-ray image. For this
image the data of the HRI observations in 1990 and 1994 have been 
combined.
The field size is 12.8\arcmin\ by 12.8\arcmin, north is up, east is left.
The image is not smoothed to show individual photons.  

Fig.\ 2 shows an overlay of contours from a
smoothed X-ray image on an optical image from the Digitized Palomar Sky Survey.
The X-ray image was smoothed with an adaptive Gaussian filter algorithm, 
using a Gaussian $\sigma$ of 17\arcsec\ for the regions of lower surface 
brightness (about 95\% of the image), and 
consecutively narrower Gaussian filters 
for higher surface brightness regions.
The lowest three contours correspond to 3, 6, and 12 $\sigma$ above background
for extended emission.

Fig.\ 3 is a overlay of the X-ray contours on the 18 cm radio map
of Clarke, Burns, \& Norman (1992).

In the following we will describe the different X-ray features 
in comparison to their counterparts in other wavelengths.

\subsection{Nucleus and jet}

Fig.\ 4 shows the background-subtracted profile of the jet 
along the jet axis. The profile was integrated in a box 30\arcsec\
wide, oriented along the jet (P.A.\ 55\arcdeg), using the long
observation in Aug.\ 1994.
The background was obtained by extracting
profiles in 30\arcsec\ wide boxes 
to either side of the jet, 
averaging, smoothing and scaling them to match
the brightness of the underlying diffuse halo southwest of the nucleus.
The shaded histogram in fig.\ 4 shows the profile before background
subtraction.

The nucleus is the bright source at the southwestern end of the jet.
The negative residuals after background subtraction
southwest of the nucleus are due to the irregular 
distribution of the absorbing material which shows up in the optical
as an irregular dust lane.
The profile of knot A may also be affected  
by this absorption. 
Judging from the absorption of the diffuse emission near the jet,
we estimate the absorption to be of the
order of 50 \% where the dust lane crosses the jet. 
The remaining uncertainty 
in the profile of knot A after background subtraction,
due to the irregularity of the dust lane,
is estimated to be on the order of 10 \% of the peak intensity.     

The knot designation follows the nomenclature of Feigelson et al.\ (1981)
and Clarke, Burns, \& Norman (1992).
In the ROSAT HRI data knots C and D as identified by Feigelson
et al.\ (1981) do not appear as individual knots.  
These knots also do not show up in the radio data,
whereas all others are clearly discernible (Fig.\ 3). 
Knots C and D were excluded from further analysis. 
Further interknot X-ray emission 
can clearly be seen between knots F and G, 
whereas there is no significant emission outside knot G, where the radio
jet blends into the inner northeastern radio lobe.

Fig.\ 5 shows the transverse profiles of the nucleus and the knots.
Superposed on each profile is a template profile obtained 
from the brightest point source in the field, 
shifted and scaled approximately to the knots' peak intensities.
The template profile is consistent with the HRI point spread function 
plus an additional artificial extent introduced by the attitude uncertainty 
of the ROSAT spacecraft 
(about 2\arcsec\ to 3\arcsec\ in this observation).

The nucleus of Cen A in X-rays is compatible with a point source. Knot A is 
marginally extended in both directions. Knot B is
clearly extended along the direction of the jet. 
Knots E, F, and G are diffuse and extended, both along and 
perpendicular to jet.  
A Kolmogorov-Smirnov test, comparing the transversal profiles of the
point source and the knots and taking into account possible errors in
central position and background subtraction, gave 
extent probabilities 
of 4, 55, 66, 96, 94 and 62 \% 
for the nucleus and knots A, B, E, F and G respectively.  

Table 1 lists the locations of the nucleus and the knots. 
The locations correspond to the intensity maxima of the knots.  
Also listed are the position angles and distances with respect to the
nucleus of Cen A.

The X-ray counts of the knots were derived by integrating the 
background-subtracted profiles shown in Fig.\ 4. 
The luminosities were calculated assuming isotropic emission, 
a distance of 3.5 Mpc, 
a power law spectrum with spectral (energy) index of $\alpha$ = 0.7
and an absorbing neutral hydrogen column density of 
$N_{\rm H}$ = $10^{21}$ cm$^{-2}$. 
Table 1 gives the luminosities in the 0.1 - 2.4 keV spectral range 
for the nucleus and for the individual knots. 

The luminosity of the nucleus in the ROSAT band (0.1-2.4 keV)  
increased significantly
between the shorter observation in July 1990 
($2.7 \pm 1.3 \cdot 10^{38}$ ergs/sec, 
assuming galactic absorption only)
and the longer one in August 1994
($6.3 \pm 0.6 \cdot 10^{38}$ ergs/sec).  
This suggests that the bulk of the X-ray emission 
originates very close to the nucleus. 
The nucleus itself is assumed to be strongly absorbed 
($N_{\rm H} \approx 10^{23} $cm$^{-2}$). 
With such a high absorption, the observable flux
would be suppressed by a factor of 500 
in the HRI energy range compared to the Galactic foreground absorption,
raising the intrinsic nuclear luminosity to more than
L$_X$ = 2-3 $\cdot 10^{41}$ ergs sec$^{-1}$ (0.1-2.4 keV) in 1994. 
However, this value for $N_{\rm H}$, 
which is frequently cited in the literature,
originates from a spectral fit to an older, 
spatially unresolved UHURU observation of Cen A (Tucker et al.\ 1973), 
and may be substantially in error. 
ROSAT PSPC spectra of the nucleus 
are contaminated by emission 
both of the nearby knot A and of the diffuse halo.
They also indicate a high intrinsic $N_{\rm H}$    
but cannot constrain its exact value. 

The luminosities of the knots within the X-ray jet of Cen A 
did not change between the 1990 and 1994 observations 
within the error margins, confirming the variability of
the nuclear source.

The spectral resolution of the HRI 
is marginally sufficient for the detection of differences 
between the amplitude channel spectra of the background and those
of the knots and the nucleus. Nucleus and knots generally 
show a harder spectrum
than the diffuse (background) emission. 
But it is not sufficient to show significant 
differences among the spectra of individual knots.
The PSPC with its better spectral resolution cannot resolve the 
individual knots in the X-ray jet. 
The necessary combination of sufficient spatial and spectral
resolution will be provided by future X-ray missions like 
AXAF or XMM.

\subsection{Southwestern X-ray feature}

Opposite to the jet, about 5.5\arcmin\ southwest of the nucleus,
an extended diffuse feature is visible in Fig.\ 1. 
Superposed on it is a bright unrelated X-ray point source which
coincides with the location of a 14 mag foreground star.
The brightest parts of the extended feature
show up as an arclike structure just outside of the southwestern
inner radio lobe in the contour plot of Fig.\ 3. 
The proximity to the outer edge of the radio lobe suggests that the 
feature consists of hot interstellar matter, 
probably from the gaseous halo of Cen A, 
which is compressed or shocked and heated by the expanding 
inner radio lobe. Similar interactions have been observed with ROSAT
in Perseus A (B\"ohringer et al.\ 1993) and Cygnus A (Carilli, Perley,
\& Harris 1994).

The polarization map of the inner radio lobes 
(Fig.\ 5 of Clarke, Burns, \& Norman 1992)
shows significant polarization 
along and aligned with the outer edge of the southwestern
radio lobe, whereas the northeastern radio lobe is less polarized
and the polarization direction is not aligned along the outer edge. 
This supports the idea of interaction between the ISM of Cen A
and the southwestern radio lobe (see discussion in
Wagner, D\"obereiner, \& Junkes 1995).
A similar feature at the outer
boundary of the northeastern lobe cannot be found.
  
After subtraction of the background and of 
the bright point source,
the X-ray feature has 660 $\pm$ 100 cts in total in the 
long HRI observation. Assuming a distance of 3.5 Mpc, a galactic
absorption column $N_{\rm H}$ = 10$^{21}$ cm$^{-2}$, 
and a Raymond-Smith spectrum with temperature 0.5 keV 
and half-solar metallicity, 
this corresponds to a total luminosity of 
$6.4 \pm 1.0 \cdot 10^{38}$ ergs sec$^{-1}$ (0.1 -- 2.4 keV). 

Imaging observations with the EXOSAT LE detector 
in the energy band 0.1 -- 2 keV have also been interpreted to
show a diffuse X-ray feature at the position of the southwestern
radio lobe 
(Morini, Anselmo, \& Molteni 1989).
However, the reported EXOSAT feature has a structure different from the 
feature detected by us in approximately the same energy band. 
Furthermore its luminosity was reported as 
$2 \cdot 10^{39}$ ergs sec$^{-1}$, 
higher by a factor of three compared to the ROSAT HRI detection. 
Since temporal variations seem unlikely, we conclude that the
EXOSAT detection was at least in part due to confusion with 
the bright X-ray point source nearby. 

\subsection{Associated point sources}

In the total HRI field-of-view 
we detect about 30 individual point sources 
with at least 5 $\sigma$, using a maximum-likelihood 
technique.
Some of the point sources in the field may be associated with
foreground objects. However, the average density of sources in the 
field is much higher than the source densities observed in similar
ROSAT observations away from the galactic plane. 
Furthermore, the source distribution is centered on Cen A.
We estimate that about 70 \% to 80 \% 
of them are associated with Cen A.
The luminosities of the detected sources are on the order of 
$10^{38}$ ergs sec$^{-1}$.
Identification of the sources is in progress.

\subsection{Diffuse emission}

Figs.\ 1 and 2 show a roughly circular halo of diffuse emission. It is 
evident from Fig.\ 2 that the absorption seen in the X-ray data   
closely matches the visual absorption indicated by the dust lane. 
Feigelson et al.\ (1981) reported the existence of two ridges of diffuse 
emission parallel to the dust lane, explained by 
emission from unresolved point sources
located in the dust and gas disk.
Our data suggest that these ridges are merely the unabsorbed edges
of the smooth diffuse halo where they adjoin the absorbed region.

The exact radial profile and total luminosity 
of the diffuse emission of Cen A has to
await a detailed background model for this observation. 
At this stage it is not possible to differentiate between a hot gaseous
halo and integrated emission of unresolved point sources. 
In particular, unresolved X-ray binaries associated with the 
gas and dust lane might contribute significantly to the 
observed diffuse emission.

However, we can give a rough estimate of
$9 \pm 3 \cdot 10^{39}$ ergs sec$^{-1}$ (0.1 - 2.4 keV, 
assuming 0.5 keV thermal emission and half-solar metallicity)
for the total luminosity of the diffuse emission 
within a radius of 12\arcmin,
corrected for the absorption by the dust lane.
Assuming that all of the diffuse emission comes from a hot gaseous halo,
Cen A is still underluminous in this respect compared to the mean (linear)    
L$_X$-L$_B$-relation for early type galaxies
(Canizares, Fabbiano, \& Trinchieri 1987).

If all the diffuse emission comes from unresolved 
discrete stellar sources 
(assuming $\alpha = 0.7$ power law spectra 
with a different counts-to-flux ratio 
in comparison to thermal emission-line spectra)
the total luminosity in the HRI energy band is calculated to 
$2 \cdot 10^{40}$ ergs sec$^{-1}$, which is twice (but compatible
within the uncertainties)  
the luminosity expected for discrete sources 
within an early-type galaxy of L$_B$ = 10.5 
(Canizares, Fabbiano \& Trinchieri 1987).

\section{Summary}

With the new X-ray data of Cen A we can improve significantly on previous 
results and clearly distinguish 
different components of soft X-ray emission.
We give luminosities and profiles 
for the nucleus and jet components (knots),
and report the detection of a new diffuse X-ray feature,
probably due to interaction of the southwestern inner radio lobe with
the hot ISM of Cen A. 

\acknowledgments

We like to thank the ROSAT team for their outstanding technical and
software support.

The ROSAT project is supported by the Bun\-des\-mi\-nis\-terium f\"ur 
Forschung und Technologie (BMFT/\-DARA) and the Max-Planck-Gesellschaft.

The Digitized Sky Survey was produced at the Space Telescope Science Institute
under U.S. Government grant NAG W-2166. The optical image is based
on photographic data obtained with the UK Schmidt Telescope.

\clearpage
\begin{deluxetable}{lccrrcrrrrr}
\tablewidth{33pc}
\tablecaption{Positions, countrates, and luminosities of the jet components}
\tablehead{
\colhead{component}  &
\colhead{P.A.}        &
\colhead{dist.}       &
\colhead{integration}  &
\colhead{total}       &
\colhead{0.1-2.4 keV luminosity}   \\
\colhead{or knot}            &
\colhead{}            &
\colhead{(")}    &
\colhead{range\tablenotemark{a}\quad(")}    &
\colhead{counts}      &
\colhead{($10^{38}$ ergs/sec)}
}
\startdata
Nucleus \tablenotemark{b}         &--&  \phantom{00}0   &
  -6.75 - \phantom{00}6.75 & 274 $\pm$ 27 & \phantom{0}6.3 $\pm$ 0.6 \nl
A                &49& \phantom{0}14   &
   6.75 -  \phantom{0}33.75 & 236 $\pm$ 36 & \phantom{0}5.4 $\pm$ 0.8 \nl
B \tablenotemark{c} &58& \phantom{0}60&
  33.75 -  \phantom{0}93.75 & 505 $\pm$ 80 & 11.6 $\pm$ 1.8 \nl
E                &56&111   &
  93.75 - 122.25 & 140 $\pm$ 30 & \phantom{0}3.2 $\pm$ 0.7 \nl
F                &56&136   &
 122.25 - 141.75 &  69 $\pm$ 21 & \phantom{0}1.6 $\pm$ 0.5 \nl
between  F and G &  &      &
 141.75 - 188.25 &  76 $\pm$ 37 & \phantom{0}1.8 $\pm$ 0.9 \nl
G                &54&204   &
 188.25 - 210.75 &  83 $\pm$ 15 & \phantom{0}1.9 $\pm$ 0.3 \nl
outside G        &  &      &
 210.75 - 282.75 &  26 $\pm$ 29 & \phantom{0}0.6 $\pm$ 0.7 \nl
\tablenotetext{a}{Integration in box along P.A.\ 55$^o$}
\tablenotetext{b}{Aug.\ 94 observation.
                  Luminosities calculated under assumption of
                  galactic absorption only}
\tablenotetext{c}{Including locations of knots C and D as observed by EINSTEIN}
\enddata
\end{deluxetable}

\clearpage

\figcaption[fig1.ps]{
ROSAT HRI image of Cen A (sum of HRI obervations in 1990 and 1994).
The field size is 12.8\arcmin\ by 12.8\arcmin, north is up, east is left.
The nucleus is marked with an arrow.
\label{fig1}}

\figcaption[fig2.ps]{
X-ray contour map (smoothed with an adaptive Gaussian filter)
overlaid on an optical image of Cen A from the Palomar Digitized 
Sky Survey. The field size is 12.8\arcmin\ by 12.8\arcmin, north is up,
east is left. The lowest three contour levels correspond to 3, 6, and
12 $\sigma$ above background.
\label{fig2}}

\figcaption[fig3.ps]{
X-ray contour map overlaid on an 18cm radio map (from Clarke, Burns,
\& Norman 1992). The field size is 12.8\arcmin\ by 12.8\arcmin, 
north is up, east is left.
\label{fig3}}

\figcaption[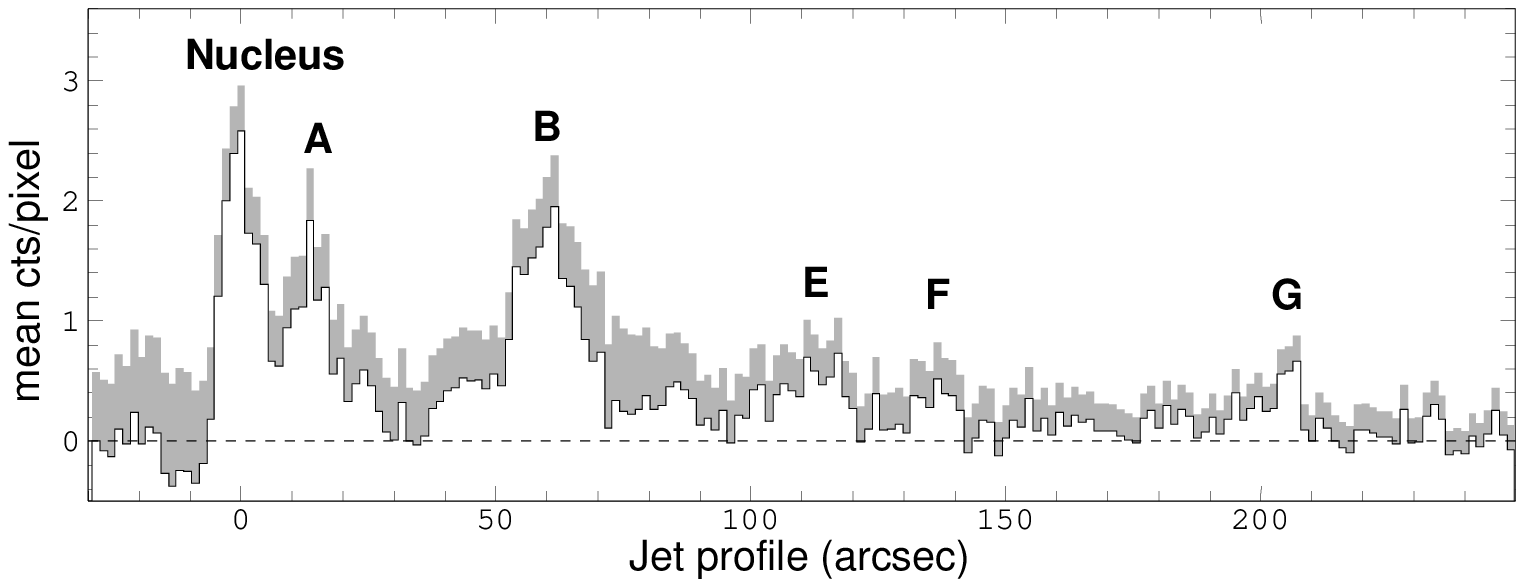]{
Background-subtracted profile of the X-ray jet (the shaded histogram 
represents the profile before background subtraction).
The nucleus and the brightest 
knots are marked according to the nomenclature of 
Feigelson et al.\ (1981). 
The bin size along the jet is 1.5\arcsec.
\label{fig4}}

\figcaption[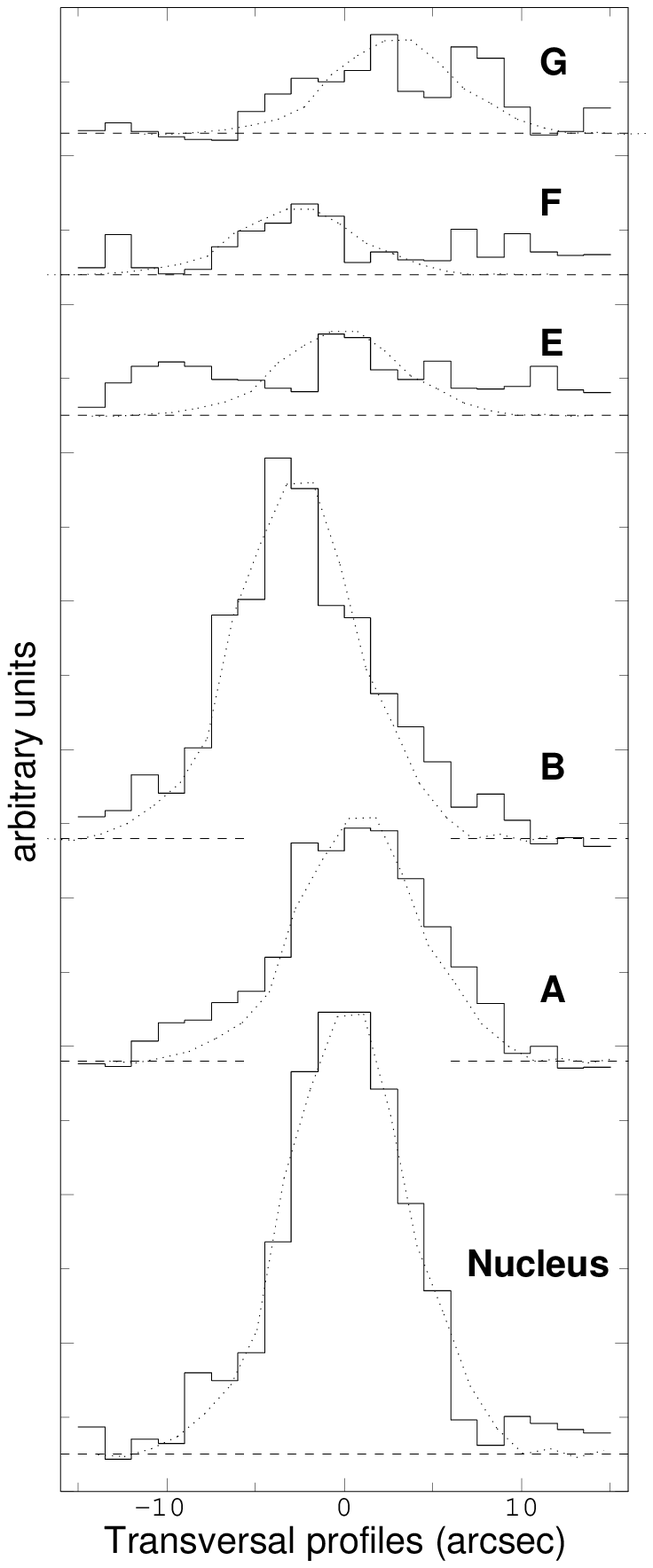]{
Background-subtracted transverse profiles of the nuclear 
source and the knots, offset along the y-axis for clarity. 
Dashed lines denote the zero levels, 
the dotted lines represent the profile of a comparison point source
scaled to the peak intensity of each source.
\label{fig5}}

\end{document}